\documentclass{llncs}
\usepackage{makeidx}  
\usepackage{graphicx}
\usepackage{amsmath}
\usepackage{mathtools}
\usepackage{url}
\usepackage{listings}
\usepackage{url}
\usepackage{multirow}
\usepackage{colortbl}
\usepackage{makecell}	 

\begin{document}

\title{On Data Flow Management: the Multilevel Analysis of Data Center Total Cost}

\author{Katarzyna Mazur\inst{1} and  Bogdan Ksiezopolski\inst{1,2}}
\authorrunning{K. Mazur and B. Ksiezopolski}

\institute{Institute of Computer Science, Maria Curie-Sklodowska University, \\ pl. M. Curie-Sklodowskiej 5, 20-031 Lublin, Poland \\
\email{katarzyna.mazur@umcs.pl}
\and
Polish-Japanese Institute of Information Technology, \\ Koszykowa 86, 02-008 Warsaw, Poland\\
\email{bogdan.ksiezopolski@acm.org}
}
\maketitle

\begin{abstract}
Information management is one of the most significant issues in nowadays data centers. Selection of appropriate software, security mechanisms and effective energy consumption management together with caring for the environment enforces a profound analysis of the considered system.  Besides these factors, financial analysis of data center maintenance is another important aspect that needs to be considered.  Data centers are mission-critical components of  all large enterprises  and frequently cost hundreds of  millions of  dollars to build, yet few  high-level executives understand the true cost of  operating  such facilities. Costs are typically spread across the IT, networking, and facilities, which makes management of  these costs and assessment of alternatives difficult. This paper deals with a research on multilevel analysis of data center management and presents  an approach to estimate the true total  costs of operating data center physical facilities, taking into account the proper management of the information flow. 
\keywords{security management, data flow management, application management, multilevel analysis, data center total cost}

\end{abstract}

\section{Introduction}

The challenges faced by companies working in nowadays complex IT environments pose the need for comprehensive and dynamic systems to cope with the information flow requirements \cite{dc1}, \cite{dc2}, \cite{dc3}. Planning can not answer all  questions: we must take a step further and discuss a model for application management. One of the possible approaches to deal with this problem, is to use the decision support system that is capable of supporting decision-making activities. In \cite{datacenter_sec}, we proposed the foundations of our decision support system for complex IT environments. Developing our framework, we examined the time, energy usage, QoP, finance and carbon dioxide emissions. Regarding financial and economic analyzes, we considered only \textit{variable} costs. However, calculating total operating cost of a data center, one needs to take into account both \textit{fixed} and \textit{variable} costs, which are affected by complex and interrelated factors. In this paper, performing  a financial analysis of security-based data flow, we present an  improved  method  for  measuring  the total  cost of its  maintenance,   exploring  trade-offs offered  by different security configurations, performance variability and economic expenditures. With the proposed approach it is possible to reduce the data center costs without compromising data security or quality of service. 

The main contributions of this paper are summarized as follows:

\begin{itemize} 
\item we enhanced previous studies on security management presented in \cite{datacenter_sec}, extending it with the analysis of fixed costs,
\item we proposed a  full cost model for data centers, being an economic method of their evaluation, in which all  costs of operating, maintaining,  and disposing are considered  important,
\item we prepared a case study, in which we:

\begin{itemize} 
\item applied the developed, financial model to an example data center, in order to show how the model actually works,
\item we evaluated the proposed economic scheme and analyzed the distribution of data center maintenance costs over five years, taking into account different security levels, and comparing them with reference to data center total costs,
\item based on the results gathered with the introduced method, we calculated possible profits and return on investment values for over five years, in order to choose the best option (considering the security of the information, along with high company incomes).
\end{itemize}

\end{itemize}

\section{Related Work}
\label{sec:realted_work}

The total cost of the data center is somewhat  elusive to project accurately. There are many subtleties which can be overlooked or are simply unaccounted for (or perhaps underestimated), over the operational life of a data center.  Looking to involve all existing elements of a modern data center into cost calculations, several approaches have been proposed in the literature (\cite{costmodel1}, \cite{costmodel2}, \cite{costmodel3}).

In \cite{costmodel1} the authors presented a way of predicting the total budget required to build a new data center. They distinguish three primary construction cost drivers, namely: power and cooling capacity and density, tier of functionality and the size of the computer room floor. They utilize proposed cost model, providing calculations for the example data center. Researchers state that their cost model is intended as a quick tool that can be applied very early in the planning cycle to accurately reflect the primary construction cost drivers. However, proposed model makes some rigid assumptions about the data center (such as the minimum floor space or the type of the utilized rack), making it quite inflexible. Moreover, as the authors themselves admit, some significant costs were not included in this cost model (for instance, the operational costs).

An interesting approach to assessing and optimizing the cost of computing at the data center level is presented in \cite{costmodel2}. Here, scientists consider five main components, that should be taken into account, while evaluating the total cost of a data center: the construction of the data center building itself, the power and cooling infrastructure, the cost of electricity to power (and cool) the servers and  the cost of managing those servers.  Besides creating a cost model for the data center, the authors examine the influence of server utilization on total cost of a data center, and state that an effective way to decrease the cost of computing is to increase server utilization. 

Another method for assessing the total cost of a data center is proposed in \cite{costmodel3}. An approach examined in \cite{costmodel3} is a part of research which seeks to understand and design next generation servers for emerging ''warehouse computing'' environments. The authors developed  cost models and evaluation metrics, including an overall metric of performance per unit total cost of ownership. They identify four key areas for improvement, and study initial solutions that provide significant benefits. Unlike our approach,  \cite{costmodel3} focuses mainly on performance and calculates data center costs only on its basis. The method proposed in our paper considers different aspects at once, which makes it flexible and suitable for heterogeneous environments.  

\section{Method  for  Estimating the Total  Cost   of a Data Center}
\label{sec:improved_method}

Although existing cost models for the data center include many components,  none of them mention security as the significant factor. However, security influences data center costs as well: proper security management translates into better utilization of central resources as well as reduced systems management and administration. 

\begin{figure}[h!p!t!]
\centering
    \includegraphics[scale=0.25]{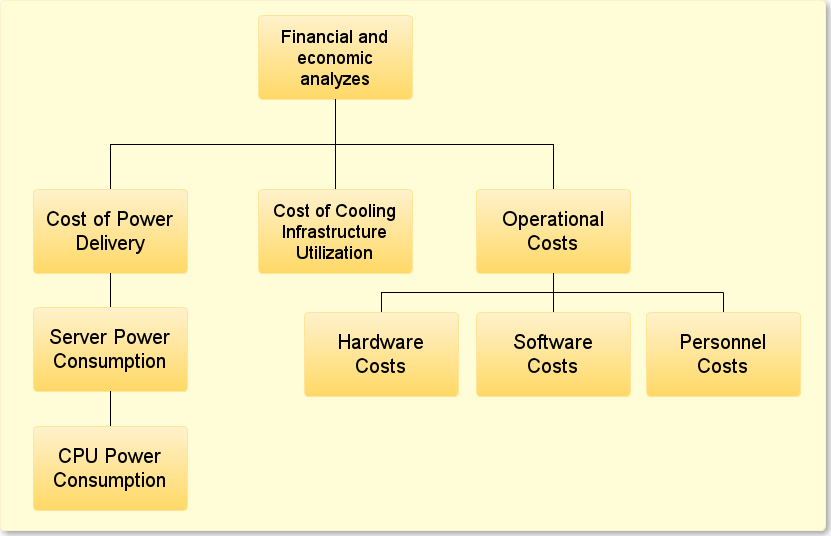}
    \caption{Components of the financial analysis.}
    \label{fig:extended_financial_analysis}
\end{figure}

In this section, we present and describe the formulas utilized in the proposed analysis process - in particular, in economic and financial analyzes, extending them with the calculation of \textit{fixed} costs. Introduced equations are used to evaluate the financial aspect of the data center maintenance.  Performing the financial analysis, we took under consideration cost of power delivery, cost of cooling, software and hardware expenditures as well as personnel salaries (both \textit{fixed} and \textit{variable} costs). Proposed method of data center cost evaluation is the sum of the present values of all costs over the lifetime  of a data center (including investment costs, capital costs, financing costs, installation costs, energy costs, operating costs, maintenance costs and security assurance costs). As shown in Figure \ref{fig:extended_financial_analysis}, in the introduced financial and economic analyzes, we distinguished $3$ main components: \textit{cost of power delivery}, \textit{cost of cooling infrastructure utilization} and \textit{operational costs}. Each of them can be further specified.

Firstly, we introduce the general formula used for calculating total energy costs.  The following equation is further detailed in terms of CPU and server utilization:  

\small
\begin{equation}
\label{eq:finance_power}
{\varsigma_{_{power}} = \kappa \cdot \chi \cdot \rho \cdot \sigma\; ,}
\end{equation}
\normalsize
 
\noindent where:

\small 
\begin{itemize}
	\item[] $\kappa$ - is the total amount of the utilized kilowatt-hours
	\item[] $\chi$ - is the total amount of hours when the server was busy 
	\item[] $\rho$ - is the total amount of days when the server was busy
	\item[] $\sigma$  - is the cost of a one kWh in US dollars
\end{itemize} 
\normalsize 

\noindent The above formula is the crucial point in computing the total cash outlay for the considered system. Original equation \eqref{eq:finance_power} is elaborated in the following sections. 

\subsection{Cost of Power Delivery}
\label{subsec:power_cost}

\subsubsection{Server Power Consumption}

The design of the electrical power system must ensure that adequate, high-quality power is provided to each machine at all times. Except the back-up system operational expenditures,  data center spends a lot of money on current power consumption - utilized  both for the compute, network and storage resources. At this point one should pay special attention to the relationship between the CPU utilization and energy consumed by other components of a working server. Since the amount of the power consumed by a single machine consists of the energy usage of all its elements, it is obvious that a decrease in the energy consumed by the CPU will result in a lower energy consumption in total. Thus, total cost of both power delivery and utilization can be summarized as follows: 

\small
\begin{equation}
\label{eq:power_delivery_cost}
\varsigma_{power} = \kappa_{busy+idle} \cdot \sigma \cdot \chi \cdot \rho\; ,
\end{equation} 
\normalsize 

\noindent where:
\small
\begin{itemize}
	\item[] $\kappa_{busy+idle}$ - is the total amount of the utilized kilowatt-hours by the server
	\item[] $\chi$ - is the total amount of hours when the server was busy 
	\item[] $\rho$ - is the total amount of days when the server was busy
	\item[] $\sigma$  - is the cost of a one kWh
\end{itemize}
\normalsize 

\subsubsection{CPU Power Consumption}

Being aware of the price of one kWh, and knowing that CPU worked $\chi$ hours through $\rho \,$days, utilizing $\kappa$ kilowatt-hours,  it is fairly straightforward to calculate the total financial cost ($\varsigma_{_{CPU}}$) of its work, using the \eqref{eq:finance_power} equation. Before we start further evaluation of the energy consumed by the CPU, we need to make some assumptions about its utilization. Let us introduce the simplified CPU utilization formula:  

\small
\begin{equation}
\label{eq:cpu_utilization_eq}
{U  =  R / C\;,}
\end{equation}
\normalsize  
\noindent where:
\small
\begin{itemize}
	\item[] $U$ - stands for  the CPU utilization, expressed in percentage
	\item[] $R$ - defines our requirements, the actual busy time of the CPU (seconds)
	\item[] $C$ - is the CPU capacity, the total time spent on analysis (seconds)
\end{itemize}
\normalsize 

\noindent Usually, the CPU utilization is measured in percentage. The requirements specified in the above formula refer to the time we require from the CPU to perform an action. This time is also known as the \textit{busy} time. CPU capacity can be expressed as the sum of the \textit{busy} and \textit{idle} time (that is, the \textit{total} time available for the CPU). Going simple, one can say that over a 1 minute interval, the CPU can provide a maximum of $60$ of its seconds (power). The CPU \textit{capacity} can then be understood as \textit{busy time} + \textit{idle time} (the time which was used plus the one which was left over). Using the above simplifications, when going multi-core, CPU capacity should be multiplied by the number of  the CPU cores ($C = C \cdot cores$). The presented equation \eqref{eq:cpu_utilization_eq} can be  further detailed as follows: 

\small
\begin{equation}
\label{eq:specific_utilization}
load \, [\%] = \frac{time_{_{session}} \, \cdot \, users}{time_{_{total}}}\; ,
\end{equation}
\normalsize 

\noindent where $time_{_{total}}$ is expressed as $time_{_{session}} \, \cdot \, users \, + \, time_{_{idle}}$.\\

Supposing a specified CPU load and assuming the server is able to handle a defined number of users within a given time, we can calculate the \textit{idle} time using the above equation. Regarding the \textit{busy} time, one should use the results obtained for the prepared, example model. This simple formula can be used for calculating the total energy consumed by the CPU. Knowing the amount of energy utilized by the CPU, it is quite straightforward to assess costs incurred for the consumed energy.  

\subsection{Cost of Cooling Infrastructure Utilization}   
\label{subsec:cooling_cost}

As the cooling infrastructure absorbs energy to fulfill its function, the cost of the cooling needs to be included in the total cost of the server maintenance.  To obtain an approximate amount of the power consumed by the cooling, one can use the equipment heat dissipation specifications, most often expressed in British Thermal Units (BTU). Specifications generally state how many BTU are generated in each hour by the individual machine. Therefore, the formula for calculating the cooling cost to keep the equipment in normal operating conditions, is given as follows (values per server):  
\small
\begin{equation}
\label{eq:cooling_power_BTU_per_hour}
\varsigma_{cooling} =  BTU_{\substack{cooling}} \cdot \sigma \cdot \chi \cdot \rho = \kappa_{cooling} \cdot \sigma \cdot \chi \cdot \rho \; ,
\end{equation} 
 \normalsize 
\noindent where:
 \small
\begin{itemize}
	\item[] $BTU_{\substack{cooling}}$ - is the amount of the BTUs generated by the cooling system 
	\item[] $\kappa_{cooling}$ - is the total amount of the utilized kilowatt-hours by the cooling system  
	\item[] $\chi$ - is the total amount of hours when the cooling system was busy 
	\item[] $\rho$ - is the total amount of days when the cooling system was busy
	\item[] $\sigma$  - is the cost of a one kWh
\end{itemize} 
\normalsize 
\subsection{Operational Costs}
\label{subsec:operational_costs}

Operational costs of the data center management depend on miscellaneous factors - among them one can enumerate salaries of the employees responsible for managing servers (along with the number of employees needed to adequately maintain and operate the data center), prices of equipment, the amount of a reduction in the value of servers with the passage of time and finally on software and licensing costs. 

\subsubsection{Hardware and Software Costs} 
\label{subsubsec:hardware_software}

In order to determine the complete cost of a data center,  software and licensing costs should be analyzed as well. The server one purchases may, or may not include an operating system.  When it comes to selecting a server OS, high-end server operating systems can be quite expensive. Except the operating system, one also needs to budget for the software applications the server will need in order to perform its tasks.  The dollar amounts can add up quite quickly in this area, depending on the role of the data center and the server itself.  It is very common in high-end server applications to offer per core licensing for some editions of the software. Dealing with hardware costs, one should be aware of server depreciation as well. Hence, the annual hardware cost is in fact an amortization cost, calculated as follows: 

\small
\begin{equation}
\varsigma_{_{\substack{hardware\\amortization}}} = \frac{\varsigma_{_{server}}}{\omega_{_{server}}}\; ,
\end{equation} 
\normalsize 
\noindent where: 
 \small
\begin{itemize}
\item[] $\varsigma_{_{server}}$ - is the purchase cost of a server (in US dollars)
\item[] $\omega_{_{server}}$ - is the average life-time of a single server (in years)
\end{itemize}
\normalsize 

\noindent When it comes to the brand new equipment, besides the amortization expenditures (the overall costs associated with installing, maintaining, upgrading and supporting a server), one should consider  purchasing costs as well. Saying so, the total, annual hardware and software cost of a single machine can be estimated using the formula below: 
\small
\begin{equation}
\varsigma_{\substack{\\hardware\\software}} = \varsigma_{_{\substack{hardware\\purchase\\cost}}} + \varsigma_{_{\substack{hardware\\amortization}}} + \varsigma_{_{\substack{total\\licensing\\costs}}} \;.
\end{equation} 
 \normalsize 
\subsubsection{Personnel Costs}
\label{subsubsec:personnel_costs}

The day may come when data centers are self-maintaining, but until then, one will need personnel to operate and maintain server rooms.  Internal personnel of the data center usually consist  of IT staff,  data center security personnel, the data center managers, facilities maintenance personnel and housekeeping personnel. If the data center contains tens - if not hundreds - of thousands of working machines, it is common to have more than one employee dealing with a given equipment. Discussing personnel costs, it can be calculated as the total number of employees  multiplied by the salary of the particular staff member (to simplify our evaluation, we can consider the average salary for every employee in the enterprise): 

\small
\begin{equation}
\label{eq:personnel_cost}
\varsigma_{_{personnel}} = (\alpha_{_{IT}} + \alpha_{_{w}} + \alpha_{_{hf}}) \cdot S_{_{avg}}\; ,
\end{equation}   
\normalsize
\noindent where:
 \small
\begin{itemize}
	\item[] $\alpha_{_{IT}}$ - is the total number of IT personnel 
	\item[] $\alpha_{_{w}}$ - is the total number of ordinary workers 
	\item[] $\alpha_{_{hf}}$ - is the total number of  the housekeeping and facilities maintenance personnel 
	\item[] $S_{_{avg}}$ - is the average salary in the enterprise (per month)
\end{itemize} 
\normalsize 

\subsection{Total Cost}
\label{subsec:total_cost}

Once the data center is built, it still requires financial investment to ensure a high-quality, competitive services with guaranteed levels of availability, protection and support continuously for 24 hours a week. Key elements in data center budgets are the power delivery system, the networking equipment and the cooling infrastructure. Besides the above most-crucial factors, there exist additional costs associated with data center operation, such as personnel and software expenses. Therefore, the real operating cost of the data center can be expressed as:  

\small
\begin{equation}
\label{eq:real_operating_cost}
{\varsigma_{total} = \varsigma_{power} + \varsigma_{cooling} + \varsigma_{operation}\; ,}
\end{equation}  
\normalsize 

\noindent where each of the defined components consists of  further operational expenditures. This concludes the discussion on calculating the total cost of a data center maintenance, resulting in the following formula (being the combination of all the above equations): 

\small
\begin{equation}
\begin{multlined}
\varsigma_{total}  = \varsigma_{power} + \varsigma_{cooling} + \varsigma_{operation}  =  \varsigma_{power} +  \varsigma_{cooling} +  \\ + \varsigma_{\substack{\\hardware\\software}}  +  \varsigma_{_{personnel}} =   \sigma \cdot \chi \cdot \rho \cdot(\kappa_{\substack{total\\power}} +  \kappa_{\substack{total\\cooling}})  + \\ + \varsigma_{_{\substack{hardware\\purchase\\cost}}}  + \varsigma_{_{\substack{hardware\\amortization}}} + \varsigma_{_{\substack{total\\licensing\\costs}}}  + \; \varsigma_{_{personnel}}  
= \sigma \cdot \chi \cdot \rho \cdot(\kappa_{\substack{total\\power}} +  \kappa_{\substack{total\\cooling}})   + \\ +  \varsigma_{_{\substack{hardware\\purchase\\cost}}} +  \frac{\varsigma_{_{server}}} {\omega_{_{server}}} +\, +  \varsigma_{_{\substack{total\\licensing\\costs}}} +    (\alpha_{_{IT}} + \alpha_{_{w}} + \alpha_{_{hf}}) \cdot S_{_{avg}} 
\,.
\end{multlined}
\end{equation} 
\normalsize

\noindent It is significant to remember that $\varsigma_{\substack{\\hardware\\software}}$ refers to the annual hardware and software costs. Therefore, when calculating the total cost of a data center maintenance, one needs to adjust this value individually, depending on the  considered analysis time interval. 

\section{Case Study: Security-Based Data Flow Management in Data Center}
\label{sec:case_study}

\subsection{Environment Definition}

To demonstrate the use of the proposed analysis scheme, we used the role-based access control approach, prepared an example data center scenario and analyzed it with the help of the introduced method. We made use of QoP-ML \cite{qopksiez}, \cite{qopbmlbook} and created by its means the role based access control model to examine the quality of chosen security mechanisms in terms of financial impact of data center maintenance.

\begin{figure}
\centering
    \includegraphics[scale=0.25]{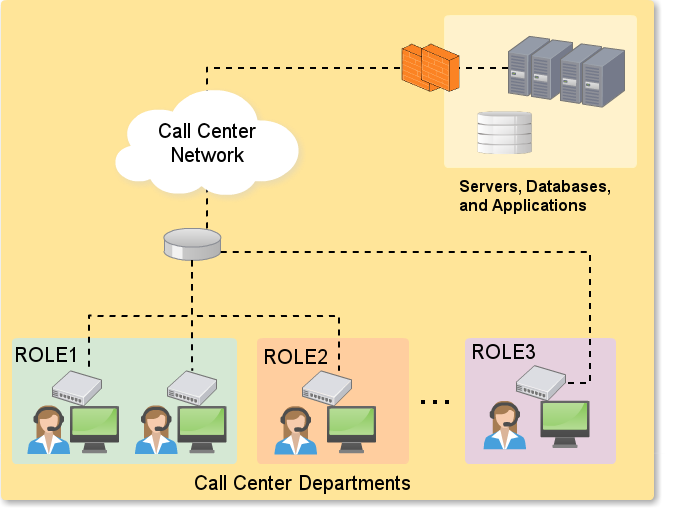}
    \caption{Example network architecture for our call center company.}
    \label{fig:callcenter}
\end{figure}

Before we perform the actual estimation of the data center maintenance cost, let us give some assumptions about the examined environment.  Consider a call center company located in Nevada, USA, managing a typical IT environment of $42$U server racks ($520$ physical servers in total, $13$ physical servers per rack). Given a specified load capacity, servers handle enterprise traffic continuously for $24$ hours. In our analysis, we assume that all the utilized applications are tunnelled by the TLS protocol.  In the considered access control method, users are assigned to specific roles, and permissions are granted to each role based on the users' job requirements. Users can be assigned any number of roles in order to conduct day-to-day tasks (Figure \ref{fig:callcenter}).

In order to emphasize and prove role's influence on data  flow management and system's performance, we prepared and analyzed a simple scenario. This scenario refers to the real business situation and possible role assignment in the actual enterprise environment. Given the example enterprise network infrastructure, consider having three roles: \textit{role1}, \textit{role2} and  \textit{role3} with corresponding security levels: \textit{low}, \textit{medium} and \textit{high}. Each server in the example call center is equipped with the Intel Xeon $X5675$ processor, being able to handle the required number of employees' connections, regardless of the assigned RBAC role.  Prepared scenarios are listed in Table \ref{table:rbac_scenarios}. QoP-ML's security models used in our case study can be downloaded from QoP-ML's project webpage \cite{qopmlweb}.  

\vspace*{-\baselineskip}
\begin{table*}[h!t!p!]
  \centering
\caption{Scenerio defined for our case study.}
\label{table:rbac_scenarios}
        \scalebox{0.55}{
        \begin{tabular}{c|c|c|c|} \cline{2-4}
 	\multicolumn{1}{c|}{} & \multicolumn{1}{c|}{\shortstack{\\\textit{role1}\\\mbox{}}} &  \multicolumn{1}{c|}{\shortstack{\\\textit{role2}\\\mbox{}}}   &  \multicolumn{1}{c|}{\shortstack{\\\textit{role3}\\\mbox{}}} \\ \hline
	\multicolumn{1}{|c|}{\shortstack{\\Access Type\\\mbox{}\\\mbox{}\\\mbox{}\\\mbox{}\\\mbox{}\\\mbox{}\\\mbox{}}} &  \shortstack{\\ E-mail, FTP, \\Web \\ Applications,\\Data Center Servers\\\mbox{}\\\mbox{}\\\mbox{}} & \shortstack{\\ E-mail, FTP, \\Web \\ Applications,\\Server S1 in DMZ, \\Server S2 in DMZ\\\mbox{}} & \shortstack{\\ E-mail, FTP,\\Web  \\ Applications\\Data Center Servers\\DMZ Production Servers\\General Office} \\ \hline 
	\multicolumn{1}{|c|}{\shortstack{\\Data Size \\(for each action separately)}} &
	
\begin{tabular}{c}
Email: 8 MB \\
FTP:  10 MB \\
Web:  3 MB \\
Applications:  4 MB \\
Data Center: 25 MB 
\end{tabular}	
	
	 & \begin{tabular}{c}
Email: 10 MB \\
FTP:  10 MB \\
Web:  4 MB \\
Applications:  6 MB \\
Server S1: 5 MB \\
Server S2: 15 MB 
\end{tabular} &

\begin{tabular}{c}
Email: 10 MB \\
FTP:  5 MB \\
Web:  4 MB \\
Applications:  6 MB \\
Data Center: 15 MB \\
DMZ Production Servers: 8 MB \\
General Office: 2 MB
\end{tabular}

 \\ \hline	
	\multicolumn{1}{|c|}{\shortstack{\\Security Mechanisms\\\mbox{}}} & \shortstack{\\TLSv1\\(RC4 MD5)} & \shortstack{\\TLSv2\\(AES/CBC 128 + SHA-1)}  & \shortstack{\\TLSv3\\(AES/CBC 256 + SHA-512)}\\ \hline
	\multicolumn{1}{|c|}{\shortstack{\\Security Level}} &  \shortstack{\textit{low}} &  \shortstack{\textit{medium}}  &   \shortstack{\textit{high}} \\ \hline
    \end{tabular}
   	}
\end{table*}

\subsection{Multilevel Assessment of Data Center Total Cost}

After introducing the environment, we present an overview of predicting the total budget required to manage an example data center, focusing on an introduced method for measuring its total cost and indicating possible gains. By the analysis of an example scenario, we try to confirm our thesis about the influence of security management on the total cost of the data center maintenance. 

\begin{enumerate}
\item \textit{Cost of Power Delivery} To calculate the total cost of power delivery, we performed both experimental and theoretical investigations. Regarding experiments, we utilized Dell UPS to measure the current power consumption of a single Dell PowerEdge R$710$ server performing operations defined in our scenario. In order to obtain the most accurate results of the CPU power consumption, we made use of our model, and performed analysis using the AQoPA tool \cite{aqopa}. To ensure the accuracy of gathered results, in our simulation, we utilized real hardware metrics, provided by the CMT \cite{cmt} for Dell PowerEdge R$710$ server. 

\begin{enumerate}
\item \textit{Server Power Consumption} According to Dell UPS in the laboratory, PowerEdge R$710$ performing defined operations  consumes on average $300$W. Since the server handles enterprise's traffic continuously for $24$ hours, its annual power consumption is equal to $2\;628$ kWhs. As stated by \url{http://www.electricitylocal.com/states/nevada/las-vegas/}, the average industrial electricity rate in Las Vegas, Nevada is $7.56$ cents $\,$ (0.0756 \$) per kWh. If the server works $365$ days a year, it will cost the company about $198.68$ \$.

\item \textit{CPU Power Consumption} When it comes to the CPU power consumption, we assumed its load to be equal to $90\%$. We performed a simulation using prepared model, considering $3$ levels of users permissions (which differ in security level).
\end{enumerate}

\noindent In Table \ref{table:cost_cpu_server} we collected power consumption costs, both for the CPU and the server in total (rounded up to the nearest dollar). As evidenced by Table \ref{table:cost_cpu_server}, considering only cost  savings related to the consumed energy, it is possible to handle about $107.64\%$ (when switching between the first and the third role) and $77.32\%$ (by changing the role from third to second) more users with the exact CPU load. Those figures, when put in the context of a large data center environment, quickly become very significant.

\begin{table*}[h!t!p!]
	\centering 
	\caption{Approximate annual cost of energy consumption for single server, rack, and whole data center  (in US dollars).}
	\label{table:cost_cpu_server}
	\scalebox{0.60}{
	\begin{tabular}{|c|c|c|c|} \hline
	 \multicolumn{4}{|c|}{ \shortstack{\\\textit{\textbf{CPU load is equal to 90\%, number of handled users varies between the roles.}}}} \\ \hline
    \diaghead{\theadfont Diag ColumnmnHead II}%
	{Server(s)}{RBAC role} & 	
	 \begin{tabular}{cc} 
    \multicolumn{2}{c}{\shortstack{\\\textit{\small role1}\\\small{(\textit{\textbf{users $\approx$ 101 361 960}})}}} \\ 
    \end{tabular} 
    & 
     \begin{tabular}{cc} 
    \multicolumn{2}{c}{ \shortstack{\\\textit{\small role2}\\\small{(\textit{\textbf{users $\approx$ 44 842 440}})}}} \\ 
    \end{tabular} 
     & 
      \begin{tabular}{cc} 
    \multicolumn{2}{c}{\shortstack{\\\textit{\small role3}\\\small{(\textit{\textbf{users $\approx$ 30 432 240}})}}} \\ 
    \end{tabular} 
     \\ \hline
   	1
   	&
   	\shortstack{\\$227 \, \$$}
   	&
   	\shortstack{\\$227    \, \$$}
   	&
    \shortstack{\\$227  \, \$$}
	\\ \hline  
	13
   	&
   	\shortstack{\\$2\;951   \, \$$}
   	&
   	\shortstack{\\$ 2\;951   \, \$$}
   	&
    \shortstack{\\$ 2\;951 \, \$$}
	\\ \hline   	
	520
   	&
   	\shortstack{\\$118\;036   \, \$$}
   	&
   	\shortstack{\\$118\;034    \, \$$}
   	&
    \shortstack{\\$118\;035  \, \$$}
	\\ \hline   
    \end{tabular}
  }
\end{table*}

\item \textit{Cost of Cooling Infrastructure Utilization} In addition to the power delivered to the compute hardware, power is also consumed by  the  cooling  resources. The load on the cooling equipment is directly proportional to the power consumed by the  compute  hardware. In such case, the cost of the cooling infrastructure is equal to the cost of the energy consumed by the server and its CPU. 

\item \textit{Operational Costs} Apart from electricity and cooling, calculating the total maintenance cost of a data center, one should take under consideration also the cost of its physical infrastructure, such as hardware amortization and the actual price of physical machines. Besides the equipment cost, operational expenditures must be covered as well. Determining the approximate, total cost of a whole data center, we assumed to use Dell Power Edge R$710$ servers ($2\;259\$$ each). However, we have include neither the network, nor the storage footprint (nor its equipment).  Regarding utilized software, we assumed that working machines have Windows Server Data Center Edition installed (whose price is equal to about $5\;497$ dollars). Except the OS, workstations use some proprietary software, which can cost about $10 \; 000$\$ on average  (\cite{ftp}, \cite{mysql}). (We assume the cost of the software for a single machine to be equal to $13\,050$ \$.)  The number of employees of the whole data center consists of security managers, system operators, call center employees and housekeeping and facility maintenance personnel, resulting in $1\;035$ employees in total. As stated by \url{http://swz.salary.com},  the median salary of security manager in US is equal to about $7\;058$ dollars per month (at the time of writing). In our estimation, we used this value as the average salary in our call center company. 
\end{enumerate}

\subsection{Discussion}

\noindent In order to prove that a proper data flow management has a significant impact on data center maintenance costs, we tried to estimate them over $5$ years. Although it might not be easily noticed at first glance, it turned out that our approach can bring meaningful savings and influence rapid \textit{return on investment} (\textit{ROI}) increase. (Table  \ref{table:r1r2compareme} explores  this concept in more detail.)  In our approach, economic profits come from the number of handled users - the more customers (served clients), the higher company profits. With the exact CPU load, the same number of working machines is capable of handling a greater number of users. By switching between the  \textit{strongest} and \textit{weakest} security mechanisms, the example call center company can achieve actual  ROI growth. Our analysis showed, that it is possible to provide effective services and  keep the utilization of hardware resources at a certain level. Since we can accomplish given goal using \textit{weaker} security mechanisms, in many situations it is wasteful to assign too many hardware resources to perform the given task. Applying the proposed solution to the existing IT environment, one can observe a serious increase in company incomes,  while preserving the efficiency, utilization and flexibility of the existing computer hardware.  

\vspace*{-\baselineskip}
\begin{table*}[h!t!p!]
    \caption{Actual profits and return on investment values for the example data center calculated for over $5$ years.}
    \label{table:r1r2compareme}
      \centering
        \scalebox{0.80}{
        \begin{tabular}{|c|c|c|c|} \cline{2-4}
     \multicolumn{1}{c|}{} & \shortstack{\\\textit{role1}} & \shortstack{\\\textit{role2}} & \shortstack{\\\textit{role3}}  \\ \hline
   	
   	\multirow{3}{*}{\shortstack{$1$st year\\\mbox{}}}  & \shortstack{Profit = $112\,541\,034\,328$\$} & \shortstack{Profit = $24\,370\,583\,132$\$} & \shortstack{Profit = $1\,890\,671\,129$\$} \\ & \shortstack{ROI = $\approx$ $247$\%} & \shortstack{ROI = $\approx$ $53$\%} & \shortstack{ROI = $\approx$ $4$\%}
	\\ \hline  

   \multirow{2}{*}{\shortstack{after $5$ years}} & \shortstack{Profit = $562\,705\,171\,640$\$} & \shortstack{Profit = $121\,852\,915\,661 $\$} & \shortstack{Profit = $9\,453\,355\,646$\$}  \\  & \shortstack{ROI = $\approx$ $1234$\%} & \shortstack{ROI = $\approx$ $267$\%} & \shortstack{ROI = $\approx$ $21$\%}

   	\\ \hline 
   	\end{tabular}
   	}
\end{table*}
\vspace*{-\baselineskip}

One of the main  pricing models of a call center company concerns \textit{cost per contact}, where all costs are combined into an unit price. The total income is then based on the number of served clients, such as calls, emails or chat sessions. For the example call center, we assumed that on average, single served customer brings the income of about $3$ US dollars. As it was proven by the time analysis presented in \cite{datacenter_sec}, server working with \textit{role3} permissions  is able to handle about $3\;474$ users within an hour having $90$\% of CPU load. Since we assumed that the number of users grows linearly, within $24$ hours, it gives us $3\;474 \cdot 24 = 83\;376$ employees a day, resulting in $83\;376 \cdot 365 = 30\; 432\; 240$ connections a year per server. If we assume that we have at our disposal the whole data center, it will turn out that we can serve roughly $30 432 240 \cdot 520 =  15 \;824\; 764\; 800$ users assigned \textit{role3} permissions a year. With the exact CPU load, using \textit{role1} permissions, server is capable of dealing with $11\;571$ users within an hour, which results in $11\;571 \cdot 24 \cdot 365 =  101 \, 361 \, 960$  served customers a year per single machine and $52 \, 708\, 219 \, 200$ clients per whole data center. 

When we translate the above calculations to the incomes and outcomes of the company, we see that the proper information management brings a variety of economic advantages. According to our  previous assumptions, given the incomes equal to $158\,124\,657\,600$\$, $69\,954\,206\,400$\$ and $47\,474\,294\,400$\$ and outcomes of $45\,584\,797\,952$\$, $45\,584\,797\,948$\$ and $45\,584\,797\,951$\$ for roles $1$, $2$ and $3$ respectively, in Table \ref{table:r1r2compareme} we calculated actual profits and ROI values of the company for over $5$ years. As it is summarized in Table \ref{table:r1r2compareme}, considering the first working year of the example call center, the efficiency of an investment is much bigger when we handle users using role's $1$ security mechanisms (comparing to the role $3$), and about $13.5$ times greater, considering roles $2$ and $3$. Real profits can be observed after the $5$ years of call center business activity. The analysis shows, that  if the company used the first role permissions instead of those from role $2$, it could gain about $5$ times more money. What is more, if we consider the third and the first role, profits grows rapidly, resulting in about $60$ times greater gain.  Since high ROI values means that the investment gains compare favorably to investment cost, and the primary goal of the call center  is a fast return on the investment, company should re-think implemented information flow mechanisms.


\section{Conclusions}
\label{sec:conclusions}

As proved by our analyzes, the main drivers of data center cost are power and cooling. In contrast to operational costs, they represent \textit{variable} costs, which vary over time and depend on many factors.  As power consumption and electricity prices  rise, energy costs are receiving more scrutiny  from senior-level executives seeking to  manage dollars.   However, focusing on the financial aspect of the data center,  one cannot forget about the proper data management. In this paper, we utilized QoP-ML to  increase company incomes, without compromising data security or quality of service. The proposed analysis scheme provides new opportunities and possibilities, not only for measuring data center costs, but also for increasing incomes. Optimization of the available computational power can be accomplished in many different ways: by modifying system configurations, switching between utilized security mechanisms, by the suitable selection of used applications and services and adequate application management.

\end{document}